\documentclass[aps,pre,twocolumn,groupedaddress,superscriptaddress]{revtex4}
\usepackage{graphicx}
\newcommand{\deffig}[4]{
\begin{figure}[tb]
  \begin{center}
  \includegraphics[width=#3 \textwidth]{#2}
  \end{center}
  \caption{ \label{fig:#1} #4}
\end{figure}
}

\begin{document}
\title{A quantum Monte Carlo algorithm for softcore boson systems}
\author{Jurij \v{S}makov}\email{jurijus@condmat.physics.kth.se}
\affiliation{
  Condensed Matter Theory, Department of Physics, Royal Institute
  of Technology, AlbaNova University Center, SE-10691 Stockholm, Sweden}
\author{Kenji Harada}\email{harada@acs.i.kyoto-u.ac.jp}
\affiliation{
  Department of Applied Analysis and Complex Dynamical Systems,
  Kyoto University, Kyoto 606-8501, Japan
}
\author{Naoki Kawashima}\email{nao@phys.metro-u.ac.jp}
\affiliation{
  Department of Physics, Tokyo Metropolitan University,
  Tokyo 192-0397, Japan
}
\date{\today}
\begin{abstract}
An efficient Quantum Monte Carlo algorithm for the simulation of
bosonic systems on a lattice in a grand canonical ensemble is proposed.
It is based on the mapping of bosonic models to the spin models in the
limit of the infinite total spin quantum number. It is demonstrated,
how this limit may be taken explicitly in the algorithm, eliminating the
systematic errors. The efficiency of the algorithm is examined 
for the non-interacting lattice boson model and compared with
the stochastic series expansion method with the heat-bath
type scattering probability of the random walker.
\end{abstract}
\maketitle

During the last few years there was an increasing number of reports on
strongly correlated quantum systems. A lot of attention has been
focused on quantum phase transitions \cite{QPT} at zero temperature,
which can be observed when parameters such as the particle
concentration and/or the interaction constants are varied. In order to
observe the quantum phase transition experimentally, one must be able
to precisely control the parameter(s), driving the transition, which
is usually very difficult in real experimental situations. Therefore,
only analytical theories and numerical simulations have been
able to provide an accurate description of the critical behavior,
associated with quantum phase transitions. Quite recently, however, a
very precise tuning of parameters was achieved in a system of
ultra-cold atoms trapped in an optical lattice, formed by the
intersection of laser beams \cite{OPT}. A transition from Mott
insulating phase to a superfluid phase was observed. It was argued
that the system is well described by the bosonic Hubbard model on a
$d$-dimensional lattice, and comparisons were made with numerical
simulations \cite{Kashurnikov,Batrouni}. This is just one example of
an experimental realization of a strongly correlated quantum system,
and a lot of experimental work will be done along these lines in the
nearest future. We believe that it is very important in
such studies to be able to provide an accurate and simple theoretical
description of the experimental system. Since the analytical solution
of the models of strongly correlated systems is usually impossible,
such a description may be in most cases provided only by the numerical
simulations.

While efficient and powerful Quantum Monte Carlo (QMC) algorithms exist
for general quantum \emph{spin} systems, the progress in the
development of the algorithms for the numerical simulations of
\emph{bosonic} systems with no hard core is much more modest. In the
present paper we describe a novel QMC algorithm, allowing efficient
simulations of the bosonic models with short-range interactions on a
lattice in the grand canonical ensemble, with average particle number
controlled by the chemical potential.




Recently an efficient QMC algorithm for the simulation of spin models
with arbitrary spin quantum number $S$ on the lattice was proposed and
implemented \cite{coarse}. It is based on \emph{coarse-graining} of the
conventional loop algorithm with split-spin representation, in which
each spin-$S$ operator is replaced by a sum of $2S$ Pauli matrices.
One update cycle of worldline configuration in this algorithm 
consists of a) placement of the vertices on the space-time lattice; 
b) creation of a pair of spin-raising or spin-lowering worms; 
c) propagation of
one of the worms through the lattice with scattering on the vertices,
resulting in changes of worldline configuration;
d) worm annihilation. The algorithm for a particular model is thus
defined by specifying a number of parameters, depending on the local
worldline configuration: density of vertices, 
scattering probabilities at vertices and the probabilities
for creation and annihilation of a pair of worms.

Holstein-Primakov (HP) transformation \cite{HP} gives a relation
between the spin systems and the boson systems.  In spin wave
theories, the transformation is used for mapping a spin problem into a
boson problem.  Here we do the opposite in order to derive a Monte
Carlo algorithm for bosonic systems from the above-mentioned one for
spin systems.  The relation can be written as $ S_i^+ = b_i^{\dagger}
(2S - b_i^{\dagger} b_i)^{1/2}, $ $ S_i^- = (2S - b_i^{\dagger}
b_i)^{1/2}b_i,$ and $S_i^z = n_i - S$, where $S_i^+, S_i^-$ and
$S^{z}_i$ are spin operators on the site $i$, and $b_i^{\dagger}$ and
$b_i$ are the boson creation and annihilation operators. At a first
glance it appears that the algorithm derived from the HP
transformation would be directly applicable only to the boson systems
that have an artificial limitation of number of particles per site
(i.e., it cannot exceed $2S$).  We show that this is not the case in
the following.

We lift the limitation by taking the limit of large $S$.  Examining
the HP transformation, we note that if $n_i$ will be kept
\emph{finite} by the chemical potential, in the large $S$ limit we
can neglect higher order terms {\it making no error}, and keep only
the lowest order in $n_i = b^{\dagger}_i b_i$ in the HP transformation
which leaves us with
\begin{equation}
\label{mapping}
b=\frac{1}{\sqrt{2S}}S^{-}\quad\textrm{and}\quad
b^{\dagger}=\frac{1}{\sqrt{2S}}S^{+}.
\end{equation}
Mapping (\ref{mapping}) allows to rewrite the Hamiltonian of a bosonic
model in terms of spin operators. Thus, if there is an algorithm for spin
systems with arbitrary $S$, and if the infinite $S$ limit of this
algorithm exists, we can easily obtain an algorithm for the bosonic
systems.

To demonstrate this idea, we consider a simple model of non-interacting
softcore bosons on a $d$-dimensional hypercubic lattice of linear size $L$
with the Hamiltonian
\begin{equation}
\label{ham_boson}
H=-\frac{t}{2}\sum_{\langle ij\rangle}(b_i^{\dagger}b_j+b_j^{\dagger}b_i)-
\mu\sum_i b_i^{\dagger}b_i,
\end{equation}
where $t$ is the (positive) hopping amplitude, $\mu$ is the chemical
potential and the first sum is over the pairs of nearest-neighbor sites. 
Using the mapping (\ref{mapping}) we can replace the bosonic operators in
(\ref{ham_boson}) with the spin operators, leading to a model
equivalent to the original bosonic model
in the limit of infinite $S$:
\begin{equation}
\label{spin-ham-final}
H=-\frac{t}{4S}\sum_{\langle ij\rangle}\left(
S_i^{+}S_j^{-}+S_j^{+}S_i^{-}\right)-
\mu\sum_iS_i^z.
\end{equation}
Since this is an $XY$ spin model, an efficient algorithm is available
for any $S$ \cite{coarse}.  Our task is, therefore, to take the
infinite $S$ limit of the algorithm. It turns out that all the
parameters defining the coarse-grained algorithm have well-defined
values in this limit as well. Below we describe the procedure of
taking this limit and give a detailed description of the softcore
boson algorithm for the non-interacting model.  Generalizations to
models with interactions, such as the on-site repulsive interaction
and short ranged repulsive and/or attractive interactions, are
straightforward. This, for instance, makes the present idea readily
applicable to the boson Hubbard model.

Naturally, boson occupation number must be positive, which leads to a
restriction on the possible values of chemical potential: $\mu < -dt$
or $|\mu| > dt$.  To apply the coarse-grained algorithm we can use the
values of parameters, derived for the general $XXZ$ model in Table I
of Ref. \onlinecite{coarse}.  Relationship between the parameters in
Ref. \onlinecite{coarse} and the parameters of our model is
\begin{eqnarray}
\label{params}
h=-\frac{|\mu|}{2dS},\,\,J=\frac{t}{2S},\,\,J'=0.
\end{eqnarray}
One has to use the results of Ref.\ \onlinecite{coarse} with caution,
since they are given for the case of positive $h$.  Therefore, in
order to use them for the present problem, we need to change the sign
of the field in (\ref{params}) and at the same time reinterpret
particle numbers denoted by $l$ and $m$ in Ref.\ \onlinecite{coarse}.
Namely, in the present article $l$ denotes the number of holes,
whereas $\bar{l}\equiv 2S-l$ denotes the number of particles. 
Accordingly, while taking the infinite $S$ limit with fixed density 
of particles we have to assume that $l$ and $m$ are
close to $2S$, whereas $\bar{l}$ and $\bar{m}$ are of order
unity.

Probability of creation of a pair of spin-raising or 
particle-number-decreasing (PND) worms in the
coarse-grained algorithm is $\bar{l}/2S$ and that of a pair of
lowering or particle-number-increasing (PNI) ones is $l/2S$. 
By taking the limit $S\rightarrow\infty$ we find, that the 
probability to create a pair of PND worms is zero. 
Corresponding probability for a pair of PNI worms is
then unity, indicating that our cycle will \emph{always} start with a
pair of PNI worms. That, however, does not mean that the number of
particles will be constantly increasing, since the worm changes
its type to the opposite one every time it changes direction as a
result of scattering on a vertex. Once the traveling worm returns
to the point of origin, it can either annihilate there, ending the
cycle, or pass through.  The probability of annihilation of a pair of
PND worms is $1/\bar{l}$ and zero for the PNI ones.

Remaining parameters, such as density of vertices and the vertex
scattering probabilities, needed for the construction of the algorithm,
can be derived by examining the values in Table I of Ref.
\onlinecite{coarse} for region IV and taking the value of $S$ to
infinity. First of all, the vertex density $B$ is given by
$B = h (lm + l\bar{m} + \bar{l}m)/2$.
To list non-zero scattering probabilities,
using the notation of Ref. \onlinecite{coarse}, we have
\begin{eqnarray}
P\left(\,\downarrow\,\left| 
      {\scriptstyle \begin{array}{cc} l & m \\ l^- & m \end{array} }
    \right. \right) & = & 2S(h-J) / (2B), \label{P0down} \\
P\left(\nearrow\left| 
      {\scriptstyle \begin{array}{cc} l & m \\ l^- & m \end{array} }
    \right. \right) & = & mJ/(2B), \label{P0diagonal} \\
P\left(\rightarrow\left| 
      {\scriptstyle \begin{array}{cc} l & m \\ l^- & m \end{array} }
    \right. \right) & = & \bar{m}J/(2B), \label{P0horizontal} \\
P\left(\nearrow\left| 
      {\scriptstyle \begin{array}{cc} l+1 & m \\ l^+ & m+1 \end{array} }
    \right. \right) & = & 1/\bar{l}, \label{P1diagonal} \\
P\left(\rightarrow\left| 
      {\scriptstyle \begin{array}{cc} l-1 & m \\ l^- & m-1 \end{array} }
    \right. \right) & = & 1/l. \label{P2horizontal}
\end{eqnarray}
Here we have set $J'=0$ and the superscript $+$ or $-$ indicates that
the type of the incoming worm is PND or PNI respectively.
Seemingly, there is a
problem with density of vertices becoming infinite in the infinite $S$
limit.  However, it should be noted that all non-trivial probabilities
of the scattering events are proportional to $1/B$, so that the
density of the scattering events remains finite.  In other words, in
the limit of infinite $S$ the situation is identical to the one that
occurs when taking the continuous imaginary time limit in a
conventional loop algorithm \cite{continuous_time}. Exploiting the
analogy to the continuous imaginary time loop algorithm, we can easily
construct a procedure for finding the time of the next scattering
event. Namely, instead of examining each vertex, it is possible to
generate the time of next event as a Poisson-distributed random number
where the average time interval or the density depends on the local
spin configuration and the type of the scattering process.

We can readily obtain the density of such events by multiplying the
scattering probabilities (\ref{P0down}), (\ref{P0diagonal}) and
(\ref{P0horizontal}) by $B$ and take the infinite $S$ limit.  Since
Eq. (\ref{P0horizontal}) yields zero, we have two non-zero scattering
densities for intervals that have no kinks in it:
\begin{eqnarray}
\Lambda\left(\,\downarrow\,\left|
{\scriptstyle \begin{array}{cc} l&m\\ l^{-}&m \end{array} }
\right.
\right) & = & \frac{|\mu|-dt}{2d},
\quad \mbox{and}\ \label{density1}\\
\Lambda\left(\nearrow\left|
{\scriptstyle \begin{array}{cc} l&m\\ l^{-}&m \end{array} }
\right.
\right)& = & \frac{t}{2}. \label{density2}
\end{eqnarray}
For the scattering probability at kinks, only the scattering probability
(\ref{P1diagonal}) will remain non-zero , since the probability
(\ref{P2horizontal}) vanishes in the infinite $S$ limit.

In order to describe the algorithm in detail, we introduce a concept
of a constant environment interval (CEI) on which the moving worm
resides. A CEI is defined as an interval ahead of the worm
in which the environment of the worm does not change in the imaginary 
time direction, i.e., 
the worldline state changes neither on the current site 
nor on any of the neighboring sites.
This interval is bounded by one of three events,
closest to the worm: a) a kink on the current site, b) a kink on one of
the neighboring sites, or c) the point of origin, where the other worm
waits for the moving worm.

Worldline configuration update cycle for the non-interacting model may be 
summarized as follows:\\[3mm]
{\bf 1)} Choose an arbitrary space-time point to place a pair of worms,
one of which will move, producing the changes in the configuration, and
another one will mark the point of origin. Always start with a PNI
(spin-lowering) worm.  Choose the arbitrary direction (up or down) for
the worm's initial movement.\\[2mm]
{\bf 2)} Determine the CEI.\\[2mm]
{\bf 3)} For each type of scattering and for each nearest neighbor site,
which is a candidate for the final scattering destination, generate 
the time of the next possible scattering event stochastically
according to the Poisson distribution with the densities
(\ref{density1}) and (\ref{density2}).\\[2mm]
{\bf 4)} If the advancement of the worm by the smallest of these times
does not take the worm out of CEI, implement the corresponding
scattering event.  In case of a back-scattering event, change the type
of the worm to the opposite one. Go back to 2.\\[2mm]
{\bf 4')} If the advancement of the worm get the worm out of CEI, advance
the worm
 to the end of the CEI.\\[2mm]
{\bf 5)} If the end point of the CEI is not a kink or the original starting
point, go back to 2.\\[2mm]
{\bf 5')} If the end point of the CEI is a kink,
attempt to scatter on it according to (\ref{P1diagonal}). 
If the scattering fails or not applicable, let the worm skip the kink
and go on. Go back to 2.\\[2mm]
{\bf 5'')} If the end point of the CEI is the original starting point,
stochastically determine, whether they will annihilate
with the probability $1/\bar{l}$ where $\bar{l}$ is the particle
number on the CEI. If it annihilates, the update cycle is terminated. 
Otherwise go back to 2.\\[3mm]
After a number of full update cycles (resulting in worm annihilation) 
the observables are measured.

To test the validity and evaluate the efficiency of the algorithm we
have performed a number of tests, comparing the results of QMC
simulation of the non-interacting boson model in three dimensions
to the exact results.
Although the model does not have any interaction terms, it is
non-trivial enough to provide us with excellent grounds for testing
because it displays Bose-Einstein condensation (BEC) and the
observables may be calculated analytically.

We have performed simulations at $k_{\rm B}T = 2t$ at 10 different
values of the chemical potential, chosen so that the resulting average
occupation number would be $n \equiv \langle N \rangle/V = 0.1, 0.2, \ldots,
1.0$. Three different system sizes are considered: $L=4,8,16$. If not
stated otherwise, for each value of system size and chemical potential
we have performed 50,000 cycles for equilibration, and another 50,000
cycles for measurement. The 50,000 measurements cycles were divided
into 10 bins of the equal length for estimating the statistical error.

\deffig{cmp_vs_occ}{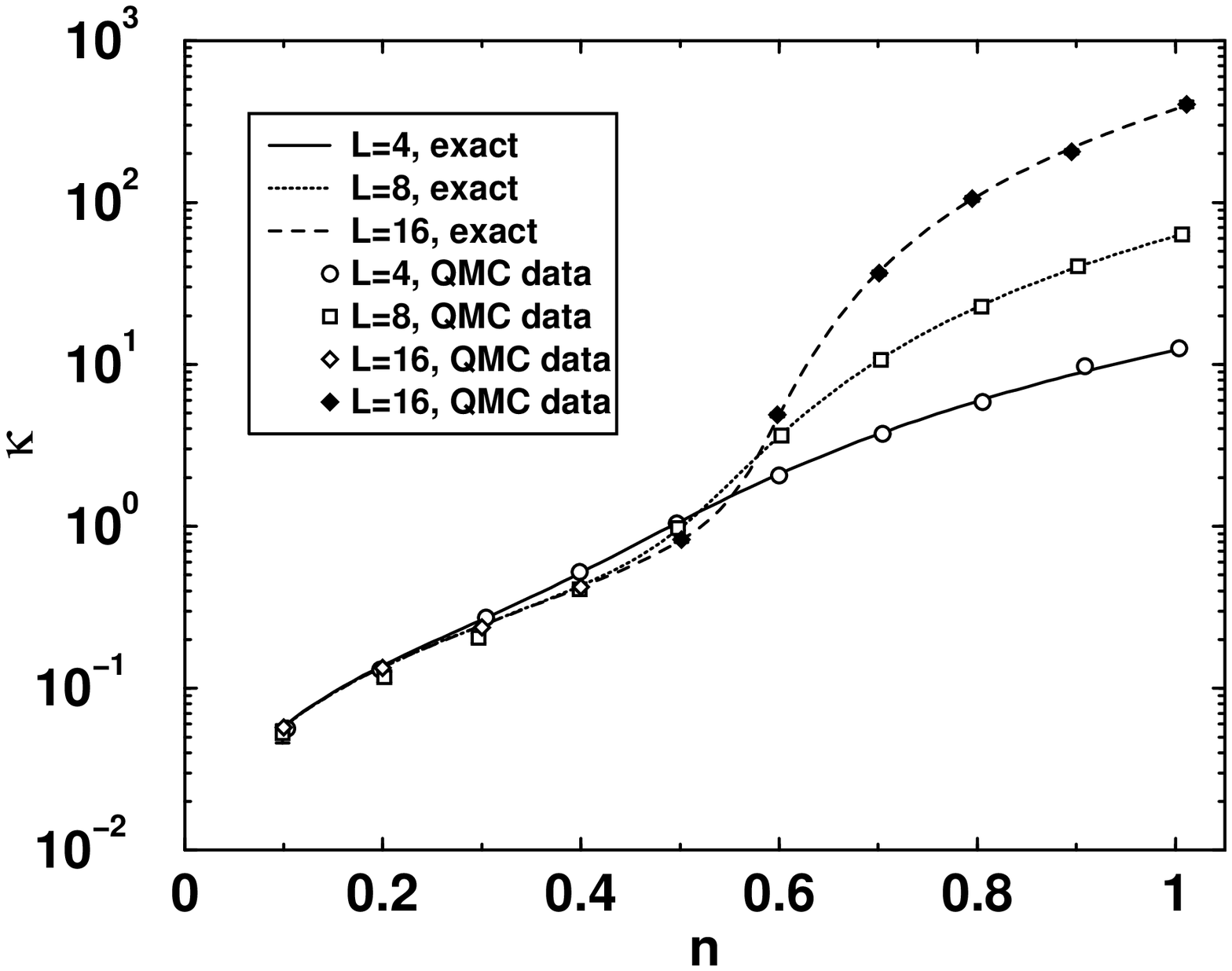}{0.48}{The compressibility plotted
against the average occupation number for three-dimensional free
lattice boson system at $k_{\rm B}T/t=2.0$. The lines are the exact
analytical values, while the symbols are the results of QMC
simulation. Uncertainty in the occupation number is in all cases
smaller than the width of the symbols. Data for $L=16$ and 
$\langle n\rangle>0.4$ is computed with the standard number of cycles 
(solid diamonds), while the data for $\langle n\rangle \le 0.4$ was
obtained with a 100 times longer simulation (open diamonds). Error bars
are in all cases smaller than the size of the symbols.}

In all cases we investigated, including cases close to criticality and
ones deep inside the superfluid phase, we found an excellent
agreement between the numerical QMC data and the exact analytical
results. As an example, Fig. \ref{fig:cmp_vs_occ} shows the dependence
of the compressibility $\kappa \equiv (\partial n /
\partial\mu)_T$ $ = (k_{\rm B}T)^{-1} (\langle N^2 \rangle -
\langle N \rangle^2) / L^d$ as a function of $n$ at a
fixed temperature $k_{\rm B} T = 2 t$. The observation of the divergent 
behavior of compressibility at BEC transition is well
within the reach of numerical simulation. For low values of
$n$ and $L=16$ we had to increase the number of
cycles 100 times in order to obtain a good statistics
because the typical lifetime of a worm becomes too short in this case.
Even after this increase, the CPU time spent for this case
is smaller than that for the superfluid cases.

In Fig. \ref{fig:super_vs_occ}, we plot the superfluid density $\rho_S$
against the average occupation number. $\rho_S$ is defined
\cite{SuperFluidDensity} as $ \rho_{S} \equiv L^{-d} k_{\rm B}T
\left[d^2F(\theta)/d\theta^2\right]_{\theta = 0} $ where $F(\theta)$
is the free energy of a system twisted by the angle $\theta$ per
lattice spacing. In QMC simulation this quantity can be measured by $
\rho_{\rm S} = L^{2-d} k_{\rm B} T \langle W_x^2 \rangle $ where $W_x$
is the sum of winding numbers of all world lines in the
$x$-direction. The possibility of measuring the winding number
fluctuation is one of the advantages of the present approach, compared
to the algorithms, such as the one used in Ref. \onlinecite{Batrouni},
that works in the fixed winding number ensemble. In
Fig. \ref{fig:super_vs_occ}, we can again see that the onset of the
condensation is captured by the QMC simulation with the present
algorithm.  
\deffig{super_vs_occ}{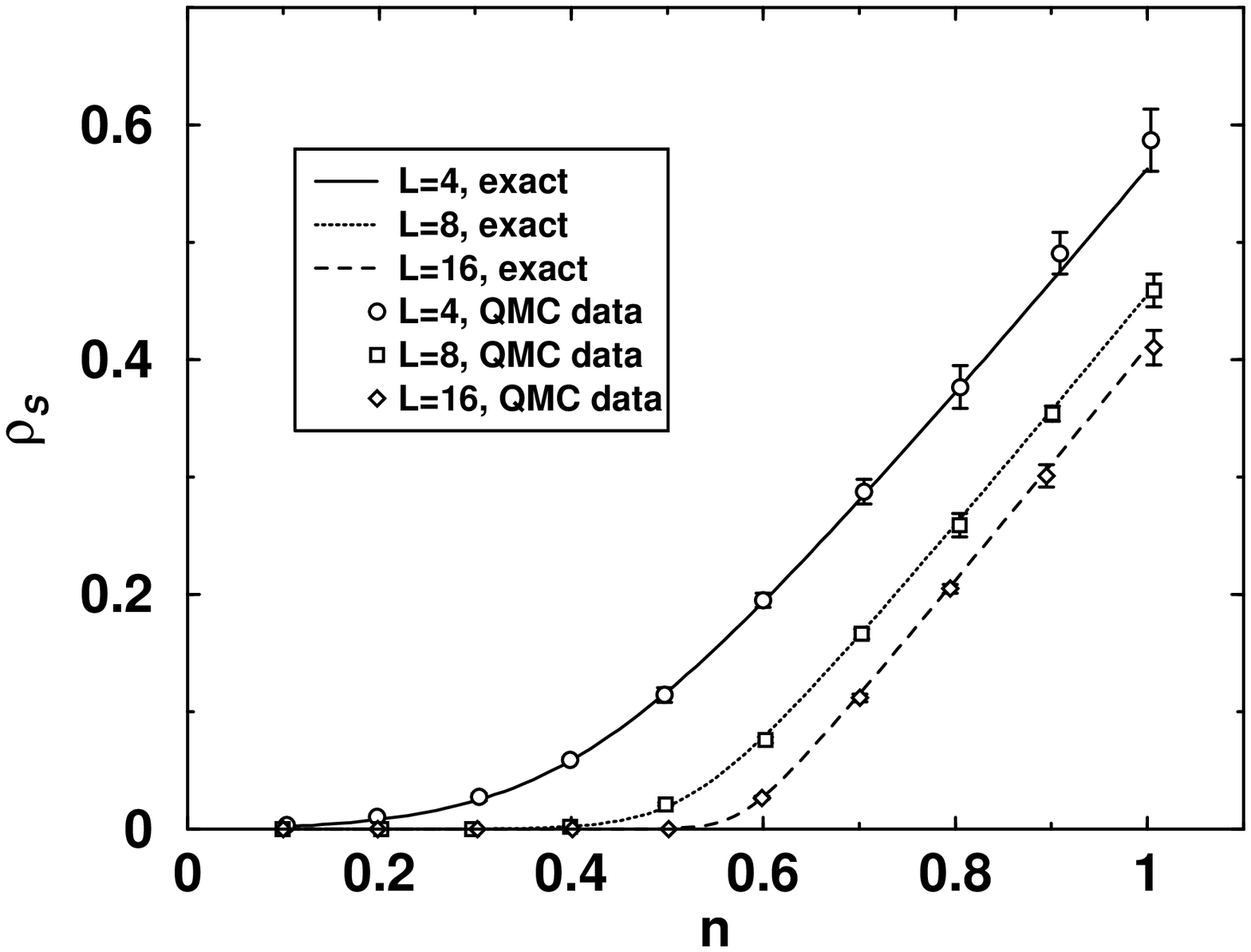}{0.48}{The superfluid density
plotted against the average occupation number for three-dimensional
free lattice boson system. Standard simulation parameters were used
for all data points. The lines are the exact analytical values while the 
symbols are the results of QMC simulation.
}

We have also compared the performance of the new algorithm with the
directed loop algorithm \cite{sas}, one of the best QMC algorithms
currently available for the simulation of softcore boson systems. The
directed loop algorithm is quite general and powerful method,
applicable, in principle, to any quantum system. However, it is up to
the user to find a set of scattering probabilities, optimizing the
efficiency of the algorithm for a given model. Due to a huge freedom
in the choice of algorithm parameters, in most cases this is a highly
nontrivial task. 
In addition, to apply the directed loop algorithm to
the softcore boson systems, one has to set an artificial upper bound
for the site occupation number. This upper bound must be taken large
enough to make the simulation free from the systematic error. 
This could be a serious disadvantage especially for bosonic system
with a random chemical potential where a large number of particles
may reside on the same site. The present algorithm is free from these
disadvantages.
For comparison purposes we have used the directed loop algorithm with a
set of simple heat-bath scattering probabilities and the site
occupation number was limited by $n_i \le 20$.

While it is hard to make a quantitative comparison of two essentially
different algorithms, we have established that our proposed algorithm
in all considered cases performs better, i.e. in all cases smaller
computational time was needed to reach the same error bars. While the
comparison of computational times does not inequivocally prove that
our algorithm is more efficient,
we note that in the vicinity of the
critical point and in the superfluid phase we have \emph{failed} 
to obtain the reliable estimates for the observables with
the directed loop algorithm within reasonable computational time. 


In summary, we have described a construction of an efficient QMC
algorithm for the simulation of softcore boson models on the lattice,
based on the coarse-grained algorithms for the spin models. By
establishing the relationship between the boson and spin operators in
the infinite $S$ (total spin quantum number) limit, we have mapped the
model of non-interacting bosons on a lattice to a spin $XY$-model in a
magnetic field. We have demonstrated, that the limit of infinite $S$
may be taken directly in the algorithm, leading to improved
performance and absence of systematic errors. The resulting algorithm
was found to perform better than existing algorithms. The result of
applications of the present algorithm to other models, such as the
Bose Hubbard model, will be reported elsewhere \cite{future}.

We are grateful to the Swedish National Allocation Committee (SNAC) and
National Supercomputer Center in Link\"oping (Sweden) for providing
computer time used for a major part of testing and simulations.
NK's work is supported by Grants-in-Aid for Scientific Research Program
(\# 14540361) from Monka-sho, Japan. J\v{S} gratefully acknowledges 
financial support from the Swedish Foundation for Strategic Research.

\end{document}